# Counts and Sizes of Galaxies in the Hubble Deep Field – South: Implications for the Next Generation Space Telescope[1]

Jonathan P. Gardner

Laboratory for Astronomy and Solar Physics, Code 681, Goddard Space Flight Center, Greenbelt MD 20771; gardner@harmony.gsfc.nasa.gov

Shobita Satyapal[2]

Space Telescope Science Institute, 3700 San Martin Drive, Baltimore MD 21218; satyapal@stars.gsfc.nasa.gov

---

*Abstract*

Science objectives for the Next Generation Space Telescope (NGST) include a large component of galaxy surveys, both imaging and spectroscopy. The Hubble Deep Field datasets include the deepest observations ever made in the ultraviolet, optical and near infrared, reaching depths comparable to that expected for NGST spectroscopy. We present the source counts, galaxy sizes and isophotal filling factors of the HDF-South images. The observed integrated galaxy counts reach >500 galaxies per square arcminute at AB<30. We extend these counts to fainter levels and further into the infrared using galaxy count models. It was determined from the HDF-N and other deep WFPC2 imaging that fainter galaxies are smaller. This trend continues to AB=29 in the high resolution HDF-S STIS image, where galaxies have a typical half-light radius of 0.1 arcseconds. We have run extensive Monte Carlo simulations of the galaxy detection in the HDF-S, and show that the small measured sizes are not due to selection effects until >29mag. We compare observed sizes in the optical and near infrared using the HDF-S NICMOS image, showing that after taking into account the different point spread functions and pixel sizes of the images, galaxies are smaller in the near infrared than they are in the optical. We analyze the isophotal filling factor of the HDF-S STIS image, and show that this image is mostly empty sky even at the limits of galaxy detection, a conclusion we expect to hold true for NGST spectroscopy. At the surface brightness limits expected for NGST imaging, however, about a quarter of the sky is occupied by the outer isophotes of AB<30 galaxies, requiring deblending to detect the faintest objects. We discuss the implications of these data on several design concepts of the NGST near-infrared spectrograph. We compare the effects of resolution and the confusion limit of various designs, as well as the multiplexing advantages of either multi-object or full-field spectroscopy. We argue that the optimal choice for NGST spectroscopy of high redshift galaxies is a multi-object spectrograph (MOS) with target selection by a micro electro mechanical system (MEMS) device. If this technology does not become available in the next few years, then the second choice would be either a mechanical MOS using movable slits or fibers, or an integral field spectrograph.

*Subject headings:*     cosmology: observations — galaxies: evolution — galaxies: statistics — instrumentation: spectrographs — space vehicles: instrumentation

## 1. Introduction

The Next Generation Space Telescope (NGST; Stockman 1997), currently in its initial design phase, is planned as a deployed ~8m telescope launched into orbit around the second Lagrangian point in 2008. The NGST Ad-Hoc Science Working Group has developed a Design Reference Mission (DRM[3]) as a set of straw-man observing projects to guide the engineering design of the observatory. The DRM is a living document, and is subject to change. A substantial part of the DRM is devoted to galaxy surveys, doing both imaging and spectroscopy. These surveys, envisioned to be of random fields, will be designed to determine the statistics of the population of galaxies in the universe. They will detect the first sources of light in the universe, as faint as 34 mag in the AB system. They will trace the assembly of galaxies and the origin of the Hubble sequence. They will measure the interaction of galaxies with their environment, and study the relation between AGN and their host galaxies. They will determine the properties of the Universe as a whole, measuring the global histories of stars, metals, and gas, and the evolution of structure in the spatial distribution of galaxies.

The Hubble Deep Field -North (HDF-N; Williams et al. 1996, Thompson et al. 1999) and -South (HDF-S; Williams et al. 2000) campaigns are the longest exposures taken to date with the Hubble Space Telescope (HST). The depth and high resolution of the images have contributed to a greatly increased understanding of galaxy formation and evolution (see, e.g., Abraham et al. 1996, Madau et al. 1996, Hudson et al. 1998, Ferguson 1998), and represent our best observational picture of galaxies in the distant universe. In this paper, we discuss the implications of the data from the two HDFs for the design of the NGST spectrograph. In section 2 we discuss the HDF source counts and use a model prediction to extend them to fainter levels and longer wavelengths. In section 3 we discuss the measured sizes of galaxies, and in section 4 we discuss the filling factor of galaxies as a function of surface brightness. In section 5 we discuss the likely redshift distribution of galaxies at the faintest levels. In section 6 we discuss several current design concepts for the NGST spectrograph. In section 7 we compare the relative multiplexing advantages of multi-object or full-field designs. In section 8 we summarize our results.

## 2. Source Counts

The deepest image made by HST, as measured in $F_\nu$, is the HDF-S STIS image (Gardner et al. 2000)[4]. This image, taken in the 50CCD clear mode, detects galaxies with reasonable completeness to fainter than 30mag in the AB system. The image was taken without a filter, and therefore the bandpass is the response function of the CCD, with sensitivity over the range 2000Å < λ < 1.1μm. The region of greatest sensitivity, however, is in the range 5000Å < λ < 1.0μm (Woodgate et al. 1998). While this region is shortward of the NGST core mission at 1.0µm < λ < 5.0µm, it provides us with an approximate number density of sources on the sky at the brightness level at which NGST will be able to do spectroscopy. The source counts in this image are plotted in Figure **1**, along with WFPC2 and NICMOS source counts. The integrated number density of objects brighter than AB = 30mag is ~$2\times10^6$ objects deg$^{-2}$, or ~5000 objects in the 3' × 3' field of view of the baseline spectrograph design.

---

[3] For more details of the NGST DRM, see http://www.ngst.stsci.edu/drm/.
[4] The HDFS data are available on the World Wide Web at: http://www.stsci.edu/ftp/science/hdfsouth/. The HDFN data are available at http://www.stsci.edu/ftp/science/hdf/hdf.html.

A more accurate way to determine the source counts in the NIR region, however, consists of using the HDF-S STIS counts and the NICMOS counts in both HDF fields to calibrate a model. The model can then be used to extrapolate the source counts to fainter magnitudes and longer wavelengths. To do this, we use the model and techniques discussed in Gardner (1998). Figure 1 is a plot of a model with the free parameters optimized to approximately fit the number counts from the HDF-S. This model, *ncmod*[5], takes a measured local luminosity function and extrapolates backwards to an assumed redshift of galaxy formation by applying a model for the luminosity evolution of galaxies. The local K-band luminosity function is taken from Gardner et al. 1997, and we apply the GISSEL96 luminosity evolution models available from Leitherer et al. 1996. The primary purpose of the model is to translate or extrapolate galaxy counts from one filter to another, or to brighter or fainter magnitudes, as we do here.

We use an open, matter-dominated cosmology with $q_0$ = 0.2 and Λ = 0. We include passive evolution of the galaxies, and extinction in the UV due to dust, but we do not include re-emission by the dust at longer wavelengths. All galaxies are considered to have formed at z=15, except for an irregular galaxy model which has constant star formation, 1 Gyr age at every redshift, and a steep faint-end slope for the luminosity function. The fit is not exact, partly due to uncertainties in the model, and partly due to systematic errors in the number counts from large-scale structure. These predictions, therefore, could be off by a factor of two. Figure 2 plots the predictions of this model for top-hat filters with a 25% bandpass, centered on 1, 3 and 5 microns wavelength.

### *3. Galaxy Sizes*

The HDF-N, other deep WFPC2 pointings (Odewahn et al. 1996), and deep NIR observations made with Keck (Bershady, Lowenthal & Koo 1998), have shown that galaxy sizes get smaller with increasing magnitude. While Ferguson & McGaugh (1995) argue that local bright galaxy surveys may be missing (or mismeasuring) a significant population of low surface brightness galaxies, their simulations show that these should be detected in the deep surveys. The HDF-S STIS 50CCD image is deeper, provides better sampling and higher resolution than the WFPC2 images, and thus can take the measurement of galaxy sizes to fainter magnitudes.

Figure 3 is a plot of the sizes of the galaxies in the HDF-S STIS image. The sizes plotted are the half-light radii, calculated by the SExtractor cataloging program (Bertin & Arnouts 1996). For Gaussian profiles, the half-light radius equals σ, while for exponential profiles, it equals 0.6 times the scale length, $r_0$. The trend that fainter galaxies are smaller is clear in this figure, and at the faintest levels the galaxies are essentially unresolved. At each magnitude, however, HST is more sensitive to compact, high surface brightness objects than it is to larger LSB galaxies, and it is necessary to understand the selection effects in this figure. We have run Monte Carlo simulations to determine our galaxy detection parameters. For each 0.5 magnitude bin, and each 0.05 arcsecond size bin, we added 20 to 40 galaxies with exponential profiles to the image, and ran SExtractor. We then computed the difference of the number counts in the modified image from the number counts in the original image. For each bin, we simulated a total of 1000 galaxies, but the number of galaxies added for each run varied to avoid introducing confusion. We constructed a completeness matrix (Drukier et al. 1988), and allowed galaxies to appear in either adjacent magnitude bin to the original, or up to ±0.75 mag from the input magnitude. In Figure 3 we plot the 80%, 50% and 20% completeness contours for our simulations. The half-light radius of the point spread function (PSF) measurements made during the HDF-S campaign gives a lower limit to the measured sizes. These measurements were of a blue star (with a similar color to the HDF-S quasar), and an unresolved redder object will have a slightly larger PSF.

---

[5] For additional information and access to the code see http://hires.gsfc.nasa.gov/~gardner/ncmod/.

We have co-added the images of the galaxies in each magnitude bin between 25 and 30. The co-added images represent an average galaxy, and have much higher signal-to-noise than any individual galaxy image. The radial profiles of these co-added images are plotted in Figure 4, and the half-light radii are marked. We have plotted these half-light radii as a function of magnitude on Figure 3 as well. Although it is clear from our simulations that selection effects are operating on the measured sizes and completeness of our catalog, it is also clear that the tendency of fainter galaxies to be more compact is a real effect. The sizes of the coadded images are smaller than they would be if the galaxies were evenly distributed over the range of sizes allowed by our simulations.

Galaxies could be either larger in the NIR than in the optical, or smaller. If the optical were typically sampling knots of star-formation, which occur within a larger, low-surface-brightness old stellar population, then we would expect sizes to be larger in the NIR. On the other hand, if the NIR were more likely to be dominated by a compact bulge, then we would expect sizes to be smaller in the NIR. Measurements of sizes of faint galaxies are complicated by the resolution of the detectors. In the HDF-S, the reduction of the NICMOS image achieved a resolution of just over 1 pixel, or 0.21". As part of the flanking field strategy, a 25ksec STIS 50CCD exposure was made of the NICMOS field, in order to measure the optical properties of the detected galaxies. This optical image has a resolution much higher than the NICMOS image, but was convolved with the NICMOS point spread function, in order to provide a fair comparison. The convolution kernel was created using Tiny Tim, (Krist & Hook 1997), and the IRAF task PSFMATCH, in which the kernel was divided in Fourier space by a Gaussian comparable to the core of the STIS PSF, (Fruchter et al. 2000). In Figure 5 we plot the half-light radii of the isolated galaxies in the summed NICMOS F110W+F160W image, compared to the half-light radii measured on the convolved and unconvolved STIS image of the field. When the STIS image is convolved to the resolution of the NICMOS image, the galaxies tend to be larger in the optical than in the NIR. On the original, unconvolved image, the galaxies are smaller. From this we conclude that galaxies tend to be more compact in the NIR than in the optical, but that the HDF-S NICMOS image is limited by its resolution.

Our simulations show that the region of Figure 3 that is populated by galaxies is not due to selection effects alone. We cannot rule out a population of large size, low-surface-brightness galaxies that would not have been detected in the HDF-S STIS image. However, for this population to exist, it could not form a continuous distribution in size with the detected galaxies. This is particularly evident in Figure 3 at about 27<AB<29 magnitude, where the 80% completeness contour is higher than most of the galaxies. At magnitudes AB>29, we can say little about the sizes of galaxies, since the completeness contours define the edge of the observed galaxy size distribution.

### *4. Isophotal Covering Factor*

The confusion limit for unresolved sources with a Euclidean distribution is typically set at 1 object in every 27 beams (Franceschini 1982). For high signal-to-noise data, number counts can be determined via an analysis of the probability of the observed fluctuations down to ~1 object per beam (Scheuer 1974). For galaxies the situation is more complex, since the size distribution, and the non-Euclidean slope of the number counts both affect the confusion noise. In this section we investigate confusion by considering the isophotal filling factor of objects in the STIS HDF-S image. We define the isophotal filling factor to be the integrated fraction of pixels, on a background subtracted image, which view a flux higher than some surface brightness. Since we are considering only images for which we have subtracted the mean background, the filling factor does not go much above 50%.

Figure 6 is a plot of the isophotal filling factor in the HDF-S STIS image. The solid line is the filling factor for the original image, and the steep rise fainter than $10^{-7}$ Jy arcsec$^{-2}$ is mainly due to noise. To determine the effects of this noise, we take three approaches. (1) For the HDF-S catalog, the object detection on the image was done by SExtractor after convolution with a Gaussian filter with full width half maximum of 3.4 pixels. We plot the filling factor of the smoothed image. This line also rises steeply due to noise at the faint end. (2) We plot the filling factor of the pixels within the segmentation map produced by SExtractor. The segmentation map consists of the contours of the detected objects at the isophote of detection. This isophote was set to be 0.65σ of the unsmoothed image, but the contours were determined on the smoothed image. It was not just a limiting isophote; there were additional criteria for the number of connected pixels applied. The segmentation map, and therefore all of the detected objects in the image at the limiting isophote, contains 5.18% of the image. (3) Finally, we constructed a noise-free image with a circular exponential profile with the same half-light radius and magnitude at the same position as the objects on the image. The profiles were clipped at 5 times the half-light radius. This noise-free image was analyzed in the same way as the real image. In Figure 6 we also plot the filling factor of the NICMOS image, where the segmentation map covers 13.44% of the image, although as discussed earlier, this higher value is probably due to the resolution of the image rather than larger sizes of galaxies.

Several things are clear from this figure. While all of the curves go to high values at the faint end, the sky is relatively empty at the detection level of the image. The HDF-S STIS image is far from confusion noise limited, and NGST spectroscopy will likewise not suffer greatly from confusion.

For NGST imaging, at ~34 mag, the situation is more complicated. The number counts plotted in Figure 2 do not predict more than one object per square arcsecond. If galaxies continue to have smaller sizes at fainter magnitudes, then NGST imaging will not be confusion limited by these faint objects. However, the "noise-free" curve in Figure 6 reaches 23% of the sky at a surface brightness level of $8\times10^{-10}$ Jy arcsec$^{-2}$, the expected $1\sigma$ surface brightness limit of a $10^5$ s exposure. This noise-free simulated image contains only those galaxies detected in the STIS HDFS image, with exponential profiles extended to $5r_{hl}$, and the isophotal covering factor at the levels below 1 nJy arcsec$^{-2}$ level is dominated by the outer isophotes of the AB~28 galaxies. This means that about one quarter of the faintest galaxies detected in deep NGST images will have to be deblended from the outer isophotes of much brighter galaxies. While galaxy detection packages such as SExtractor are designed to do this, it will impact both the design of the observations, and the depth and completeness of the results. For example, in order to use confused data to derive flat fields, Arendt, Fixsen & Moseley (2000) show that it would be necessary to dither the observations on all scales up to the field of view of the detector, (see also Fixsen, Moseley & Arendt 2000), a capability which will need to be built into the observatory.

## 5. The Redshift Distribution of Galaxies

The techniques of determining photometric redshifts from broadband colors have achieved prominence in recent years. Recent interest has been motivated by studies of the HDF-N which found and characterized populations of $2.0<z<3.5$ and $3.5<z<4.5$ actively star-forming galaxies, (U- and B- band dropouts) (Madau et al. 1996), and Keck spectroscopic confirmation of the technique in the HDF-N and in ground-based surveys (Lowenthal et al. 1997; Steidel et al. 1996). Despite the successes, care must be taken not to over-interpret photometric redshifts, particularly of individual objects. The technique works best for galaxies with strong spectral breaks, such as the 4000Å break at low redshift, and the Lyman break or strong absorption by the Lyman$\alpha$ forest at high redshift. The effects of dust on photometric redshifts have not been well determined. Nonetheless, as an indicator of general trends in the population of galaxies, photometric redshifts can extend our understanding to fainter levels than the spectroscopic ability of ground-based telescopes. For our purposes, photometric redshifts can help us to determine the advantages of using pre-selection on the basis of photometry for doing an NGST spectroscopic redshift survey. These advantages, primarily for a multi-object spectrograph (MOS), come when the total number of galaxies per field of view is greater than the possible number of slits in the instrument. Pre-selection of the targets of interest via photometric redshifts (e.g. only galaxies with $z>2$) would reduce the number of objects for which spectroscopy would be necessary.

Figure 7 is a plot of redshift distribution of galaxies with $25<I_{814}<28$ in the HDF-N, as determined by Fernandez-Soto, Lanzetta & Yahil (1999). One quarter of the galaxies are at $z>2$. A spectroscopic redshift survey, using a MOS, and designed to study high-z galaxies would first find the objects using broadband imaging. The imaging could be done in several filters, and photometric redshifts could be used to select the targets of interest. If we assume that the number of slits available in a single MOS exposure is of order 1000, and if there are ~5,000 galaxies per 3'×3' field of view, then 5 configurations of the MOS (or more, depending on clustering and confusion) would be needed to get redshifts for all of the galaxies. However, nearly all of the $z>2$ objects could be studied spectroscopically in just 2 configurations of the MOS. The selection of NGST spectroscopic targets by photometric redshifts is more powerful than it might appear at first. If we assume that the deep spectroscopic surveys will be done in the same region as the deep imaging surveys, then broadband colors with very high signal-to-noise ratio would be available. One of the limiting factors in the accuracy of photometric redshifts is photometric error. The low-redshift, bright galaxies, which are of limited interest in an NGST redshift survey, would be those for which the photometric redshifts are most accurate. Contamination by low-redshift but intrinsically very faint galaxies would be relatively low, and these objects might be of enough interest to study spectroscopically in any case. We make a more detailed comparison of instrument concepts in the next section.

The disadvantages of photometric redshift selection in advance of a MOS survey are twofold. First, inaccurate photometric redshifts could scatter objects into or out of the selection criteria. When low-redshift galaxies are mistaken for high-redshift galaxies, it wastes the spectroscopic observing time. When high-redshift galaxies scatter into the low-redshift bins, it is more serious, since this contamination would not be easily detected. Simulations show, however, that this effect can be reduced with sufficiently high signal-to-noise in the photometric observations and sufficiently accurate galaxy templates, (Teplitz et al. 2000b), or by using principal component analysis with a large training set of spectroscopically determined redshifts, (Connolly & Szalay 1999). An additional disadvantage of photometric redshift selection for a MOS in comparison to full three-dimensional spectroscopy is that it reduces the probability of the serendipitous discovery of pure line-emission sources. To search for such objects with a MOS would require configurations studying "blank" sky, or objects detected photometrically at very faint levels.

## 6. Possible NGST Spectrograph Designs

NGST will gather unprecedented spectroscopic and photometric data on galaxies out to the highest redshifts. As we have shown, the anticipated number counts in the NGST field of view is large, expected to be close to ~5000 at AB=30 mag, requiring an efficient method for multiplexed spectroscopy. There are several possible instrument concepts for carrying out spectroscopy over a large field of view. These include imaging Fourier Transform Spectrometers, (Graham et al. 1999), dispersive-based multi-object spectrometers (MacKenty et al. 1999; Moseley et al. 1999), image slicers (LeFevre et al. 1999), and tunable filters or Fabry-Perots (Satyapal et al. 1999). In this section, we investigate quantitatively the sensitivity trades between these three fundamental instrument concepts.

### 6.1 Dispersive Spectrographs

The sensitivity of a long slit grating-based spectrometer is given by:

$$SNR = \frac{S_\lambda \frac{\lambda}{R} \frac{t}{N}}{\sqrt{\left[\left(N_R^2 + I_{dark}\frac{t}{N}\right)npix_{psf}(\lambda)\right] + \left(S_\lambda \frac{\lambda}{R}\frac{t}{N}\right) + B_\lambda \frac{\lambda}{R} npix_{psf}(\lambda)}} \quad 1.$$

Where $S_\lambda$ and $B_\lambda$ are the source and background photon rates including instrument throughput and grating efficiency factors, which are strong functions of wavelength for a given blaze angle. Here $\lambda$ is the wavelength, $R$ is the spectral resolution, $N$ is the number of grating settings or gratings required to obtain the entire spectrum, $t$ is the total integration time, $I_{dark}$ is the dark current, and $npix_{psf}(\lambda)$ is the number of pixels in the point spread function. We assume that the maximum exposure time per readout is limited by the cosmic ray rate to 1000 seconds, and $N_R$ is the total effective read noise from the multiple 1000 second exposures required at each grating setting. The grating band average efficiency is assumed to be 50%. Slit losses are not included in these calculations.

### 6.2 Imaging Fourier Transform Spectrometers

A Fourier Transform Spectrometer (FTS) acquires a full 2 dimensional image of the entire field of view observed. Successive images corresponding to different positions of a moving mirror are acquired and the Fourier transform of the resulting interferogram yields a spectrum of every pixel in the array. The signal to noise ratio in this case is given by the expression:

$$SNR = \frac{\frac{\sqrt{N}}{M}\int S_\nu d\nu \frac{t}{N}}{\sqrt{\left[2\left(N_R^2 + I_{dark}\frac{t}{N}\right)npix_{psf}(\lambda)\right] + \left(\int S_\nu d\nu \frac{t}{N}\right) + \int B_\nu d\nu\, npix_{psf}(\lambda)}} \quad 2.$$

Here, $S_\nu$ is the signal photocurrent including instrument efficiency factors (assumed to be constant with frequency in this analysis), $B_\nu$ is the background photocurrent, $N_R$, $I_{dark}$, $t$, and $npix_{psf}(\lambda)$ are as above, while $N$ is the number of samples in the interferogram, and $M$ is the number of spectral channels required for the desired spectral resolution. (This derivation is given by Bennett 1999a). The factor of two in the detector noise term arises for dual port interferometers, where all photons are collected using two focal plane arrays. The spectral resolution of an FTS is determined by the maximum scan length of the moving mirror and is constant with respect to frequency. For a Nyquist sampled spectrum, $N$ is twice the spectral resolution. As can be seen from this equation, an FTS views all photons within its pass band simultaneously. In photon noise limited operation, and in the absence of band reducing filters, the detectors are also subject to the associated shot noise from the entire pass band. The assumptions used for all calculations in this paper are summarized in Table 1.

### 6.3 Imaging with Tunable Filters

As in the case of the FTS-based spectrograph, an imaging tunable filter-based spectrograph views the entire field at once. The full spectrum is obtained by imaging the field with successive filters, or by tuning a Fabry-Perot (FP). Here the signal to noise ratio is given by equation 1, with the factor $\lambda/R$ replaced by $\delta\lambda$, the bandwidth of the filter in wavelength. $N$ now represents the number of filters or FP settings required to obtain the entire Nyquist-sampled spectrum. As can be seen, the detectors in an FP-based system receive reduced signal compared with the FTS case since only those photons from the source within the bandwidth of the filter at each setting are passed. However, the associated shot noise is also reduced compared with the FTS case. When obtaining a spectrum covering the full wavelength range under background-limited operation, and assuming identical instrument throughputs, the sensitivity of an FTS and an FP are equal.

### 6.4 Point Source Sensitivities and Spatial Resolution

In Figure 8, we plot the single object point source sensitivity as a function of spectral resolution to obtain the full 1-5 µm spectrum with an FTS, a grating (for the detector size stated in Table 1), and an FP. The flux density is given at 3 µm for a signal to noise ratio of 10. As can be seen, the grating system is the most sensitive for a single object.

The large aperture of NGST provides for the capability of high spatial resolution imaging. Adequate sampling of the point-spread function implies pixel scales of 30 mas. However, for obtaining the highest sensitivity for integrated galaxy spectra, the optimum choice of plate scale for the wide-field spectroscopic survey work is a function of the light distribution of faint galaxies at high redshift, the zodiacal background, and the detector noise. Using the radial light profiles plotted in Figure 4, we show, in Figure 9, the signal to noise ratio as a function of pixel size to detect an AB=28 mag galaxy in $10^5$ seconds of total integration time. For the scientific goal of maximizing the detection rate of high z galaxies, the optimum pixel scale is between 100 and 200 mas.

### 6.5 Multi-Object Spectroscopy

While the sensitivity of an FTS (or FP) is not competitive for point source detections, as the number of sources in a given field of view is increased, dispersive-based spectrometers are limited by the number of simultaneous spectra that can be obtained in a single exposure with a given detector array size. In Figure 10, we include the impact of the anticipated number counts in the NGST field on observing efficiency for the three instrument choices. Specifically, the integration time required to obtain a 10 σ detection on 10 nJy sources in the entire 3' x 3' FOV is plotted verses spectral resolution. The predicted integrated number counts in the NGST spectrograph FOV at 3 μm is ~$4 \times 10^3$. At R=100, a multiple slit spectrograph can obtain spectra of all sources with a 4k × 4k detector. At R=1,000, we assume that the spectrum of approximately 1,000 sources can be obtained at once. The total time to achieve the stated sensitivity at 10 nJy is calculated. We have not included in this calculation the added overhead in target pre-selection. In addition, in observing with an FTS, a simultaneous deep broad-band image is obtained with each spectral data cube. Observations with a multi-object spectrograph require initial imaging observations for target positions. As can be seen from Figure 10, at higher resolution, the efficiency of the grating-based spectrograph in obtaining full spectra of the NGST galaxy field is most evident.

## *7. MOS vs. IFS*

When the scientific goal of a project is to get spectroscopy with full wavelength coverage of every spatial point in an area larger than the field of view, and at spectral resolutions where the detector noise is significantly lower than the shot noise from the background, the FTS, FP, IFS and MOS all take the same exposure time to make the observation, assuming identical throughput factors. When the goals are more modest, then the different spectrograph concepts have varying strengths and weaknesses. The FTS offers the greatest flexibility in spectral resolution, and could be considered a replacement for the NGST camera as well as the spectrograph. At spectral resolutions where the dispersive spectrographs are detector noise limited, the FTS offers greater area or sensitivity in the same exposure time than the dispersive spectrographs used in mapping mode (Bennett 1999b). However, when one does not need spectra of every point on the sky, the FTS does not achieve the ultimate sensitivity on individual targets or small areas. Narrow wavelength coverage over a wide field of view is best done with a FP, but it is inefficient to get high spectral resolution over a wide range of wavelength. The greatest sensitivity for individual sources is obtained by dispersing the background (as well as the signal) in a grating-based spectrograph. To disperse the background while multiplexing the observations requires either a target selection device, such as a MOS, or an integral field spectrograph (IFS) such as an image slicer (Weitzel et al. 1996; Le Fevre et al. 1999).

Both a MOS and an IFS re-format the observed sky, pass the light through a dispersive spectrograph, and project it onto a detector array. Ignoring slit losses, both achieve the same sensitivity. Assuming the same pixel scale and detector size, and an infinitely configurable MOS, both observe the same total area on the sky. The MOS selects this area from within a larger field of view, while the IFS observes this area as a contiguous region. Alternative designs conceptually similar to the IFS include long slits or picket fence spectrographs. The advantages of a particular design depends on the geometry of the sources one wishes to observe; a MOS favors widely scattered compact sources, while an IFS favors individual extended objects. With these assumptions, the choice is clear. Since the prime NGST science goals include observing the high redshift universe – made up of widely scattered, compact galaxies – a MOS will make the most efficient use of the observing time.

The assumption of an infinitely configurable MOS is not valid, however. A MOS made with a micro electro mechanical systems (MEMS) technology will come the closest, while a macroscopic mechanical MOS selector (e.g. Crampton et al. 1999) could place severe constraints on the observing efficiency, but would require less technology development. While the mechanical MOS would be able to observe target galaxies when they are sufficiently numerous, it would likely not be able to observe closely packed galaxies, or be able to build up spectral imaging data cubes without considerable effort. The slit size would be fixed, making it inefficient at observing complex regions. This departure from the ideal must be compared to the advantages and technical simplicity of an IFS.

To make this comparison we apply the number counts and sizes presented above to spectral resolutions R=100, and R=1000. We take $10^5$ seconds per spectral octave as the exposure time, reaching $K_{AB} \sim 27.4$ at R=1000, $5\sigma$ for continuum point sources. For emission line sources, the limit is $5 \times 10^{-20}$ ergs/cm$^2$/s. The z=2.73 gravitationally lensed galaxy, MS1512-cB58 has been observed in the near-infrared by Teplitz et al. (2000a), and has rest-frame equivalent widths of 106Å in Hα, 26Å in Hβ, 97Å in [OIII] λ5007Å, and 37Å in [OII] λ3727Å. To obtain at least $5\sigma$ in all of these lines, a galaxy at z=3.4, (i.e., with Hβ in the K band), as faint as $K_{AB} \sim 29.3$ in the continuum could be observed. For continuum detections at R=100, the limit is $K_{AB} \sim 29.6$. From the integral model plotted in Figure 2, there are about 550 galaxies per square arcminute at these depths. If we take 25% of the objects to be "interesting" by some criteria (i.e., photometric redshifts with z>2), then we would wish to observe about 130 galaxies per square arcminute. It is important to note here that we do not believe that the selection of galaxies with photometric redshifts z>2 is the only criterion that will be used. That will depend on the particular science questions that are being asked. Alternative selection possibilities include slices in apparent magnitude, color selection, or morphological selection. It is likely, however, that some selection will be desired (if the alternative is to spend extra exposure time to get every object in the field of view), and therefore we take 25% selection as a typical value.

Each octave of the spectrum in an IFS must be allocated 1.33R pixels, where R is the spectral resolution. Likewise, this is the footprint of a spectrum on the MOS detectors. If we assume that a 4096 × 4096 detector array is available, and that the spatial sampling is 0.1 arcsec per pixel, then the IFS would have a field of view equal to 35/R arcmin$^2$. This corresponds to a field 35" × 35" at R=100, or 11" × 11" at R=1000. If we assumed a MOS equal in size to the detector, then packing the spectra with 2.5 spatial pixels each (on average), and allowing 1.3R pixels of the detector as wasted space on the sides, the theoretical limit for number of slit positions that the MOS could observe would be (4096-1.3R) × 1640/1.3R, or 48,000 for R=100, or 3400 for R=1000. This theoretical limit assumes that all parts of the sky are equally interesting. We have shown, however, that most of the sky is empty at the depths reachable by NGST spectroscopy, so we have to make a correction for the positions of the galaxies. The number of independent slit areas can be determined by dividing these numbers by 4 so that the galaxy could appear anywhere in the area, and still produce a full, non-overlapping spectrum. Simple simulations have shown that when the number of independent slit areas is equal to the number of targets, then approximately 60% of the targets can be observed with non-overlapping spectra. When the numbers are not equal, it quickly becomes possible to observe >90% of either the number of slits or the number of targets, whichever is smaller. Assigning targets relative priorities, and using artificial intelligence algorithms to optimize slit placement, would result in a more efficient observing strategy.

So far, we have considered a MOS array equal to the detector array size. If we limit ourselves to a 2048 × 2048 MOS array, one would lose a factor of 4 in the number of independent slit areas, but would gain some back some of that area by no longer having any edge effects. In addition, the use of a dichroic to give two octaves of the spectrum would return a factor of 2. With an even smaller 1024 × 1024 MOS, one might use a larger MOS slit scale of 0.2". Therefore, a $2048^2$ MOS with 0.1" shutters or mirrors, and with a dichroic, would give 6300 independent slit areas at R=100, or 630 at R=1000. These slit areas would be selected from a field of view of 3.4' × 3.4'. The MOS would be able to observe all detectable galaxies at R=100, or about half of the detectable z>2 emission-line galaxies at R=1000 in each $10^5$ second exposure.

In contrast, an IFS targeting 130 "interesting" objects per square arcminute would be able to observe 44 objects per field of view at R=100, and 4 to 5 targets at R=1000, making it difficult to observe a representative sample of the universe. A mechanical MOS, with ~100 slits or fibers and potentially severe geometric constraints would return a comparable number of objects at R=100. Even at R=1000, the advantage of a mechanical MOS over an IFS is not strongly evident, particularly if the technical simplicity of an IFS were to translate into a greater investment in detectors, or if a larger spatial pixel scale, and thus a greater field of view, were used. Since the mechanical MOS would not be efficient at observations of targets other than high redshift galaxies, the second choice for the NGST spectrograph is not clear. If a MEMS MOS can be developed on a suitable schedule, and within the NGST cost constraints, then that would be the best choice to conduct the NGST core science.

## *8. Summary*

The Next Generation Space Telescope will be constructed to image, obtain redshifts for, and make a detailed study of galaxies in the high redshift universe. Existing HST observations of the Hubble Deep Fields show that the faintest galaxies are numerous, but small in size, and do not reach the confusion limit with the spatial resolution available to HST in the optical, or NGST in the near infrared. The HDF datasets reach depths comparable to that expected for NGST spectroscopy, and we have discussed the implications of these data for the design of the NGST spectrograph. Using these data to constrain a model of the galaxy counts, we have extended the counts into the near-infrared to give an idea of the number of targets that will be available for spectroscopy. An analysis of the sizes of the galaxies shows that in the deep, high-resolution HDF-S STIS image, the typical half-light radius of the galaxies is 0.1 arcseconds. The sizes in the near infrared are likely to be comparable or slightly smaller. An analysis of the isophotal filling factor of galaxies shows that at the limits of galaxy detection in the HDF, HST imaging is not confusion limited. Of the options for the NGST spectrograph, including FTS, FPs, MEMS and mechanical MOS's or IFSs, the best choice for observing the high redshift universe is the MEMS MOS. If the MEMS MOS turns out to not be technically feasible, then the second choice, between the mechanical MOS and the IFS, is not as clear.

We would like to acknowledge useful discussions with Rick Arendt, Santiago Arribas, Charles Bennett, Mark Dickinson, Harry Ferguson, Bob Fosbury, James Graham, Matt Greenhouse, Randy Kimble, Simon Lilly, Knox Long, John MacKenty, Harvey Moseley, Marcia Rieke, Eric Smith, Massimo Stiavelli, Dan Watson and Bruce Woodgate. We wish to thank John Mather, Peter Stockman and the NGST ASWG for the design reference mission. We wish to thank Bob Williams and the HDF-N and HDF-S teams for making available the data used in this paper.

## *References*

Abraham, R. G., Tanvir, N. R., Santiago, B. X., Ellis, R. S., Glazebrook, K., & van den Bergh, S. 1996, MNRAS, 279, L47
Arendt, R. G., Fixsen, D. J., & Moseley, S. H. 2000, ApJ, in press, astro-ph/0002258
Bershady, M. A., Lowenthal, J. D., & Koo, D. C. 1998, ApJ, 505, 50
Bennett, C. L. 1999a, to appear in, "Imaging the Universe in Three Dimensions: Astrophysics with Advance Multi-Wavelength Imaging Devices," eds. W. van Breugel & J. Bland-Hawthorn, astro-ph/9908245.
Bennett, C. L. 1999b, to appear in, "NGST Science and Technology Exposition", eds. E. P. Smith & K. S. Long
Bertin, E., & Arnouts, S. 1996, A&AS, 117, 393
Connolly, A. J., & Szalay, A. S. 1999, AJ, 117, 2052
Crampton, D., et al., 1999, to appear in, "NGST Science and Technology Exposition", eds. E. P. Smith & K. S. Long
Drukier, G. A., Fahlman, G. G., Richer, H. B., & VandenBerg, D. A. 1988, AJ, 95, 1415
Ferguson, H. C., & McGaugh, S. S. 1995, ApJ, 440, 470
Ferguson, H. C. 1998, Rev. Mod. Astr. 11, 83
Fernandez-Soto, A., Lanzetta, K. M., & Yahill, A. 1999, ApJ, 513, 34
Fixsen, D. J., Moseley, S. H., & Arendt, R. G. 2000, ApJS, in press, astro-ph/0002260
Franceschini, A. 1982, Astrophys. & Space Science, 86, 3
Fruchter, A. et al. 2000, in preparation
Gardner, J. P., Sharples, R. M., Frenk, C. S., & Carrasco, B. E. 1997, ApJ, 480, L99
Gardner, J. P. 1998, PASP, 110, 291
Gardner, J. P. et al. 2000, AJ, 119, 486
Graham, J. 1999, to appear in, "NGST Science and Technology Exposition", eds. E. P. Smith & K. S. Long, astro-ph/9910442
Hudson, M. J., Gwyn, S. D. J., Dahle, H., & Kaiser, N. 1998, ApJ, 503, 531
Kelsall, T., Weiland, J. L., Franz, B. A., Reach, W. T., Arendt, R. G., Dwek, E., Freudenreich, H. T., Hauser, M. G., Moseley, S. H., Odegard, N. P., Silverberg, R. F., & Wright, E. L. 1998, ApJ, 508, 44
Krist, J. E., & Hook, R. N. 1997, in, "The 1997 HST Calibration Workshop, With a New Generation of Instruments", eds. S. Casertano, R. Jedrzejewski, T. Keyes, & M. Stevens, (STScI: Baltimore), p. 192
Le Fevre, O., et al. 1999, to appear in, "NGST Science and Technology Exposition", eds. E. P. Smith & K. S. Long
Leitherer, C. et al. 1996, PASP, 108, 996
Lowenthal, J. D., Koo, D. C., Guzman, R., Gallego, J., Phillips, A. C., Faber, S. M., Vogt, N. P., Illingworth, G. D., & Gronwall, C. 1997, ApJ, 481, 673
Madau, P., Ferguson, H. C., Dickinson, M. E., Giavalisco, M., Steidel, C. C. & Fruchter, A. 1996, MNRAS, 283, 1388
MacKenty, J. W., et al., 1999, to appear in, "NGST Science and Technology Exposition", eds. E. P. Smith & K. S. Long
Moseley, S. H. et al. 1999, to appear in, "NGST Science and Technology Exposition", eds. E. P. Smith & K. S. Long
Odewahn, S. C., Windhorst, R. A., Driver, S. P., & Keel, W. C. 1996, ApJ, 472, L13

Satyapal, S., et al., 1999, to appear in, "NGST Science and Technology Exposition", eds. E. P. Smith & K. S. Long
Scheuer, P. A. G. 1974, MNRAS, 166, 329
Steidel, C. C., Giavalisco, M., Pettini, M., Dickinson, M., & Adelberger, K. L. 1996, ApJ, 462, L17
Stockman, H. S. 1997, ed., ``Next Generation Space Telescope: Visiting a Time When Galaxies Were Young'', (A. U. R. A., Inc.: Baltimore).
Teplitz, H. I., McLean, I. S., Becklin, E. E., Figer, D. F., Gilbert, A. M., Graham, J. R., Larking, J. E., Levenson, N. A., & Wilcox, M. K. 2000a, ApJ, submitted
Teplitz, H. I., et al. 2000b, in preparation
Thompson, R. I., Storrie-Lombardi, L. J., Weymann, R. J., Rieke, M. J., Schneider, G., Stobie, E., & Lytle, D. 1999, AJ, 117, 17
Weitzel, L., Krabbe, A., Kroker, H., Thatte, N., Tacconi-Garman, L. E., Cameron, M., & Genzel, R. 1996, A&AS, 119, 531
Williams, R. E., et al. 1996, AJ, 112,1335
Williams, R. E., et al. 2000, in preparation
Woodgate, B. E., et al., 1998, PASP, 110, 1183

Table 1

Assumptions for signal to noise calculations

| | |
|---|---|
| Telescope diameter | 8 m |
| Detector size | 4kx4k |
| Plate scale | 0.03"/pixel (critically sampled for diffraction limited performance at 2 µm) |
| Detector dark current | 0.02 e/s |
| Detector read noise | 3 e⁻ |
| Detector quantum efficiency | 0.90 |
| Wavelength coverage | 1-5 µm |
| Background | Zodiacal background model from DIRBE (Kelsall et al. 1998) |
| Maximum single exposure time | 1,000 s (determined by cosmic ray rate) |

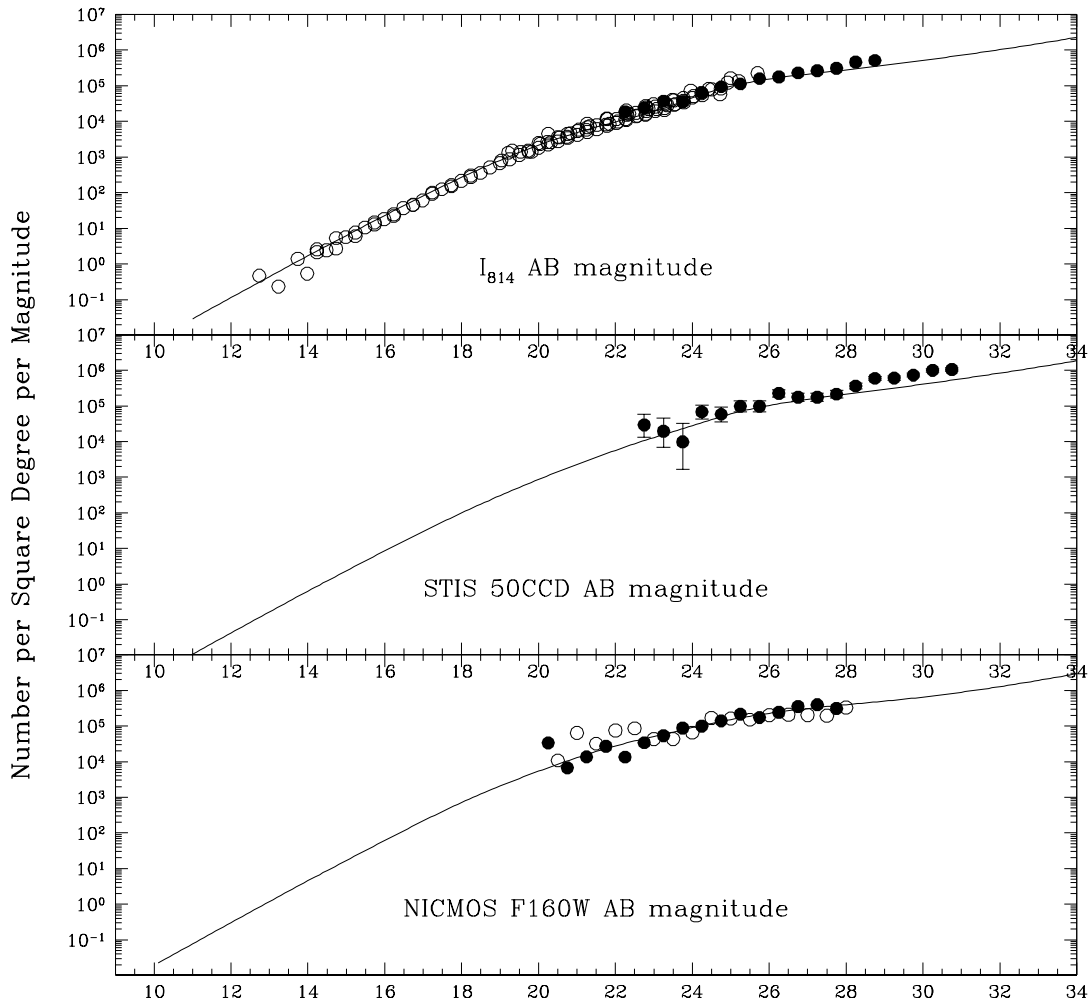

Figure 1. Number counts in the HDF-S images, in WFPC2 $I_{814}$, STIS 50CCD and NICMOS F160W filters. We also plot a model in which the free parameters have been set to fit the data, (Gardner 1998). In the top panel, the open symbols are ground-based I-band counts converted to the $I_{814}$ system, and the filled symbols are the HDF-N counts from Williams et al. (1996). In the bottom panel, the filled symbols are the NICMOS F160W galaxy counts from the HDF-S; the open symbols are the HDF-N counts from Thompson et al. 1999.

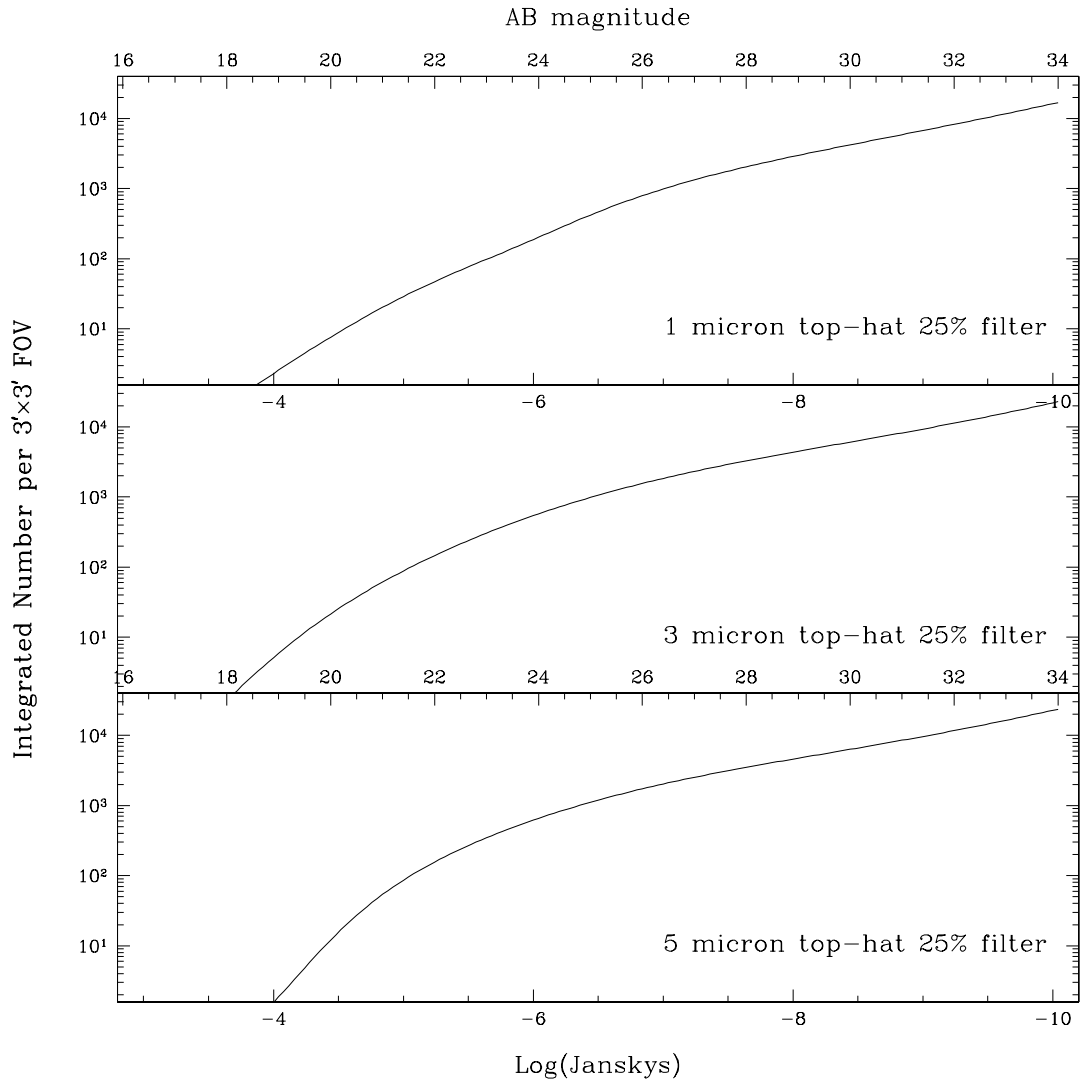

Figure 2. Model predictions for the integrated number of galaxies in 3 NGST filters as a function of flux in Janskys. The filters are 25% bandwidth ideal top hat filters centered at 1.0, 3.0 and 5.0 microns. The field of view is assumed to be 3'×3'.

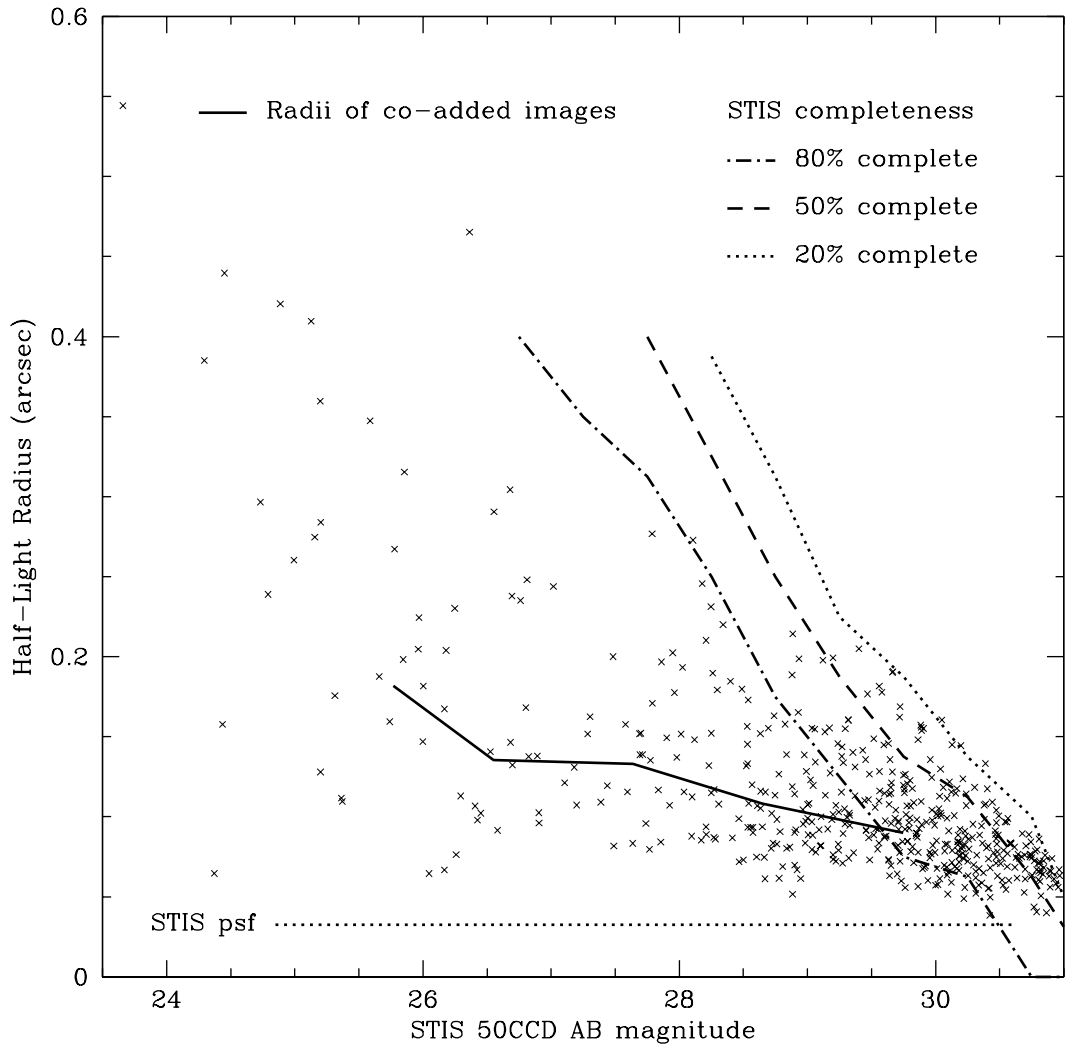

Figure 3. Sizes of objects on the HDF-S STIS 50CCD image. We plot the half-light radii as a function of AB magnitude. The high resolution and faint limiting magnitude of the STIS image confirms and extends the trend seen in deep WFPC2 images, where fainter galaxies are smaller. We have conducted extensive Monte Carlo simulations to determine the selection effects operating on this Figure, and we plot the 80%, 50% and 20% completeness limits. The half-light radius of the STIS point-spread function is plotted as a line representing our measurement lower limit, although this has a weak dependence on color. We also plot the half-light radii of the binned and co-added data from the next Figure.

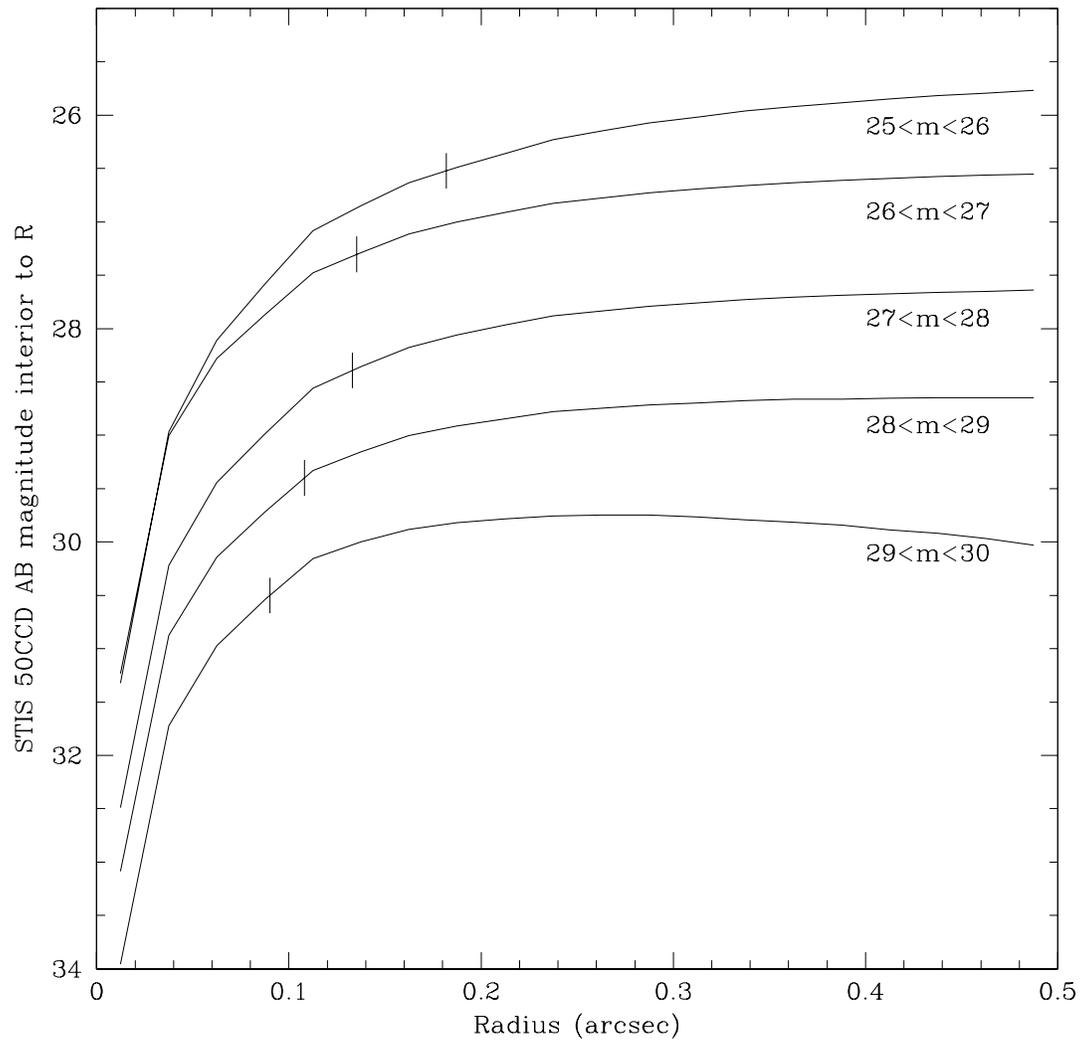

Figure 4. Radial profiles of co-added images. The tick marks are placed at the half-light radii.

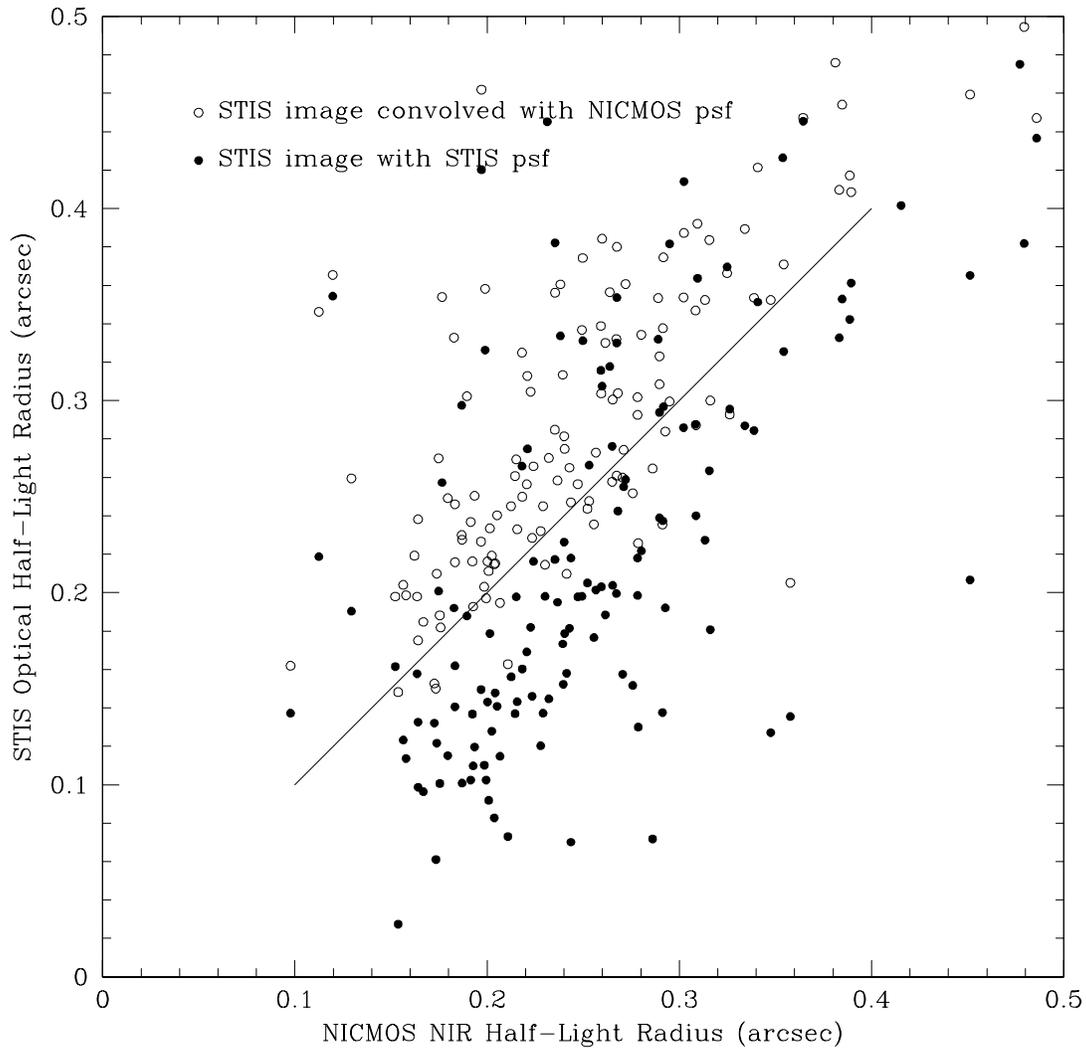

Figure 5. Sizes of galaxies in the HDF-S NICMOS F160W (H-band) image. As part of the flanking field strategy, a 25ksec STIS 50CCD exposure was made of the NICMOS field. This figure compares the half-light radii of the objects in the optical vs the NIR. When the optical image is convolved with the point spread function of the NIR image, the measured sizes are slightly larger in the optical. However, using the full resolution of the STIS image, the sizes are smaller. Galaxies are more compact in the NIR than in the optical, but the HDF-S NICMOS image is limited by the resolution.

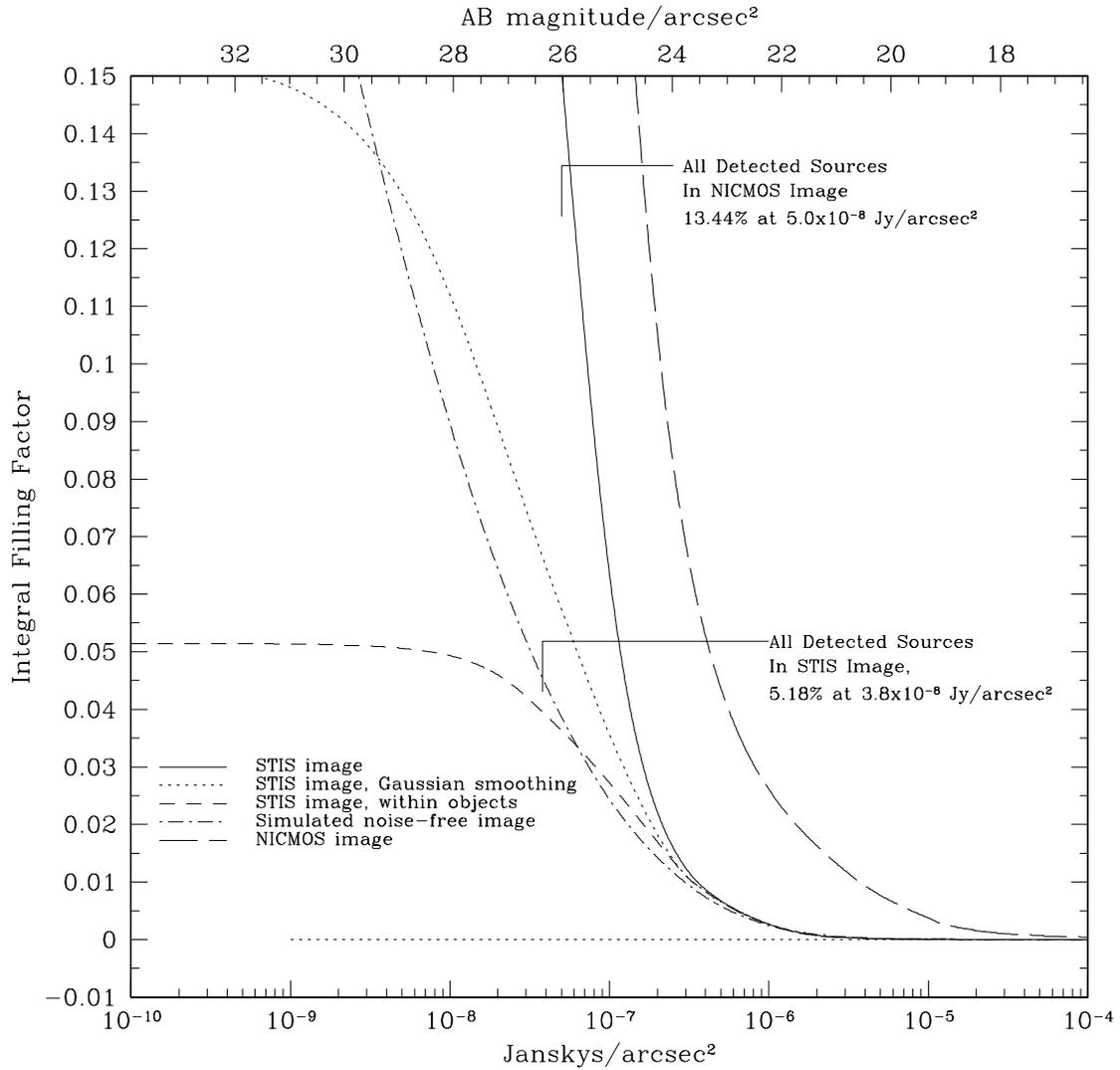

Figure 6. Filling factor of objects on the HDF-S STIS 50CCD image. The QSO and bright star (and their diffraction spikes) have been masked. The isophotal detection limit used by SExtractor to construct the catalog is $0.65\sigma$. All detected objects cover a total of 5.18% of the image at this isophote (after convolution with a 3.4 pixel Gaussian filter).

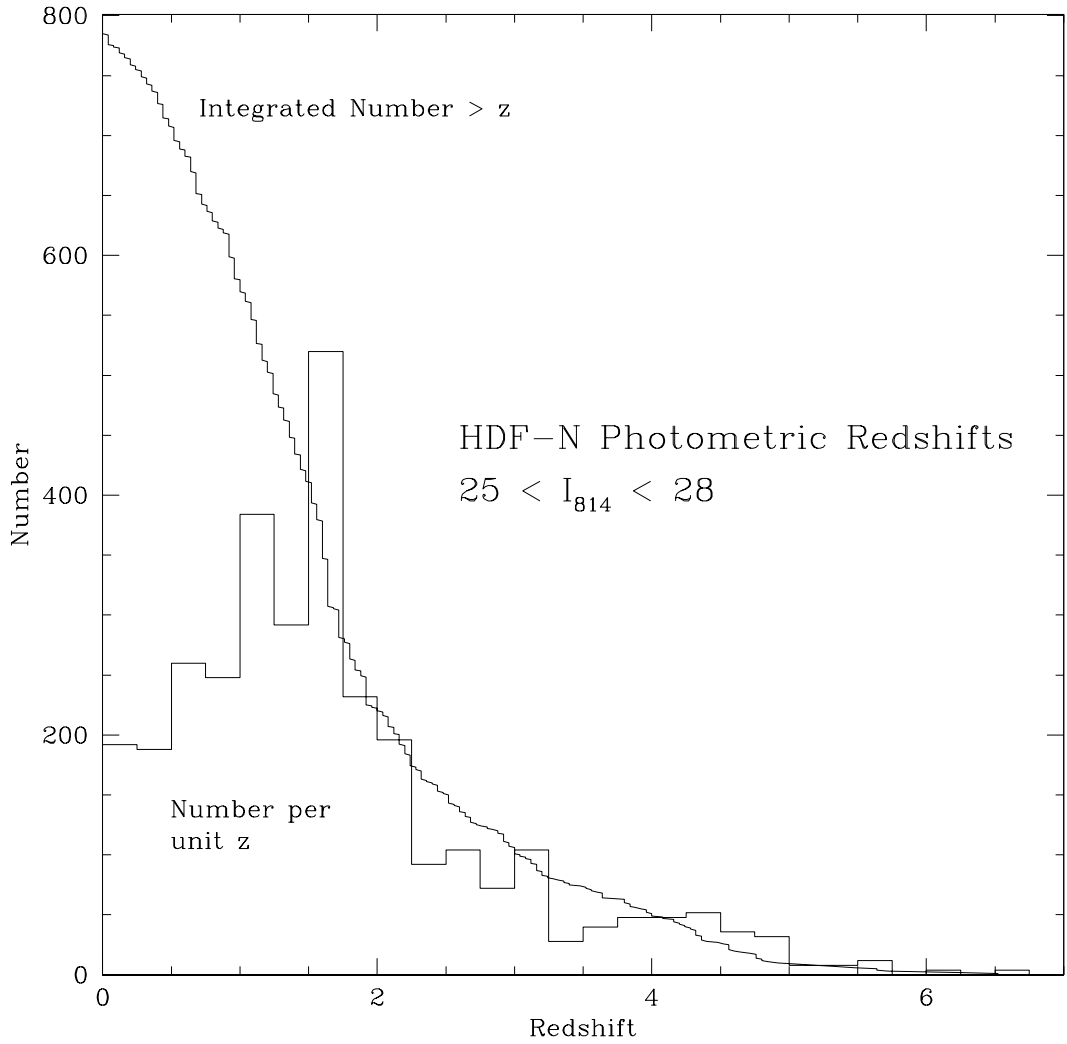

Figure 7. Photometric redshifts of galaxies with $25<I_{814}<28$ in the HDF-N, as determined by Fernandez-Soto et al. (1999). We plot both the differential histogram of the redshifts, and the integrated number greater than z. This figure shows that using photometric redshifts to preselect objects with z>2 would reduce the number of target objects by a factor of four.

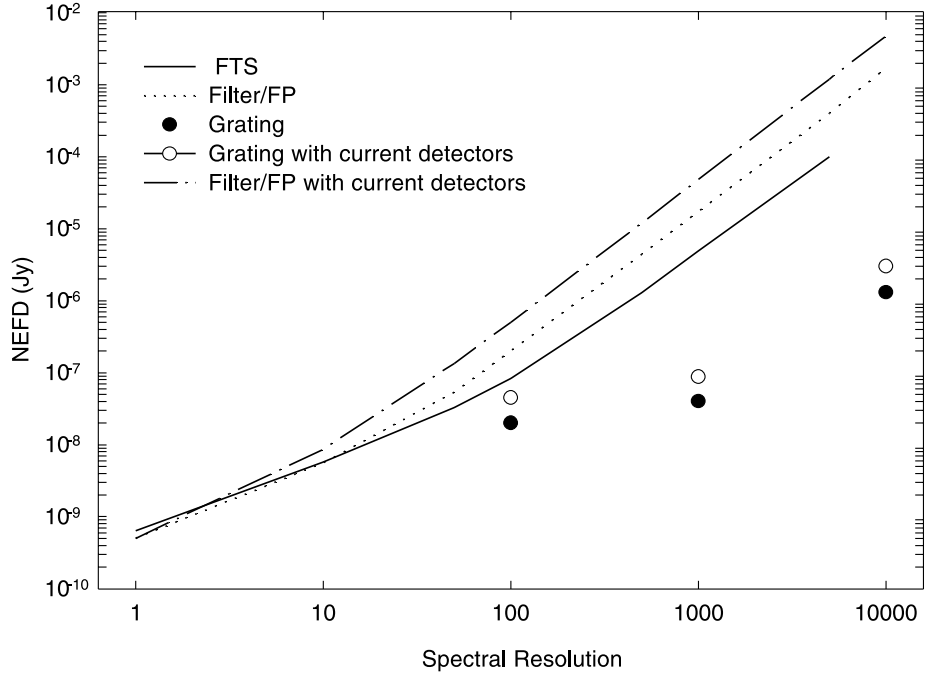

Figure 8. Point source sensitivity to obtain full spectrum for an FTS, FP, and grating vs. spectral resolution. Flux density is given for a signal to noise ratio of 10 at 3 um and a total integration time available to obtain the whole spectrum of $10^5$ seconds. As can be seen, highest sensitivity to detect a single point source is achieved by a grating-based system. This effect is most dramatic at high resolution. At low resolution, when the noise is dominated by the background shot noise in the Fabry-Perot/Filter case, the sensitivity of the FP/filter system is equal to that of the FTS. For the Fabry-Perot system, as the spectral resolution is increased, the background flux reaching the detector is reduced and the noise becomes dominated by detector noise. At these resolutions, the FTS is more sensitive than an FP.

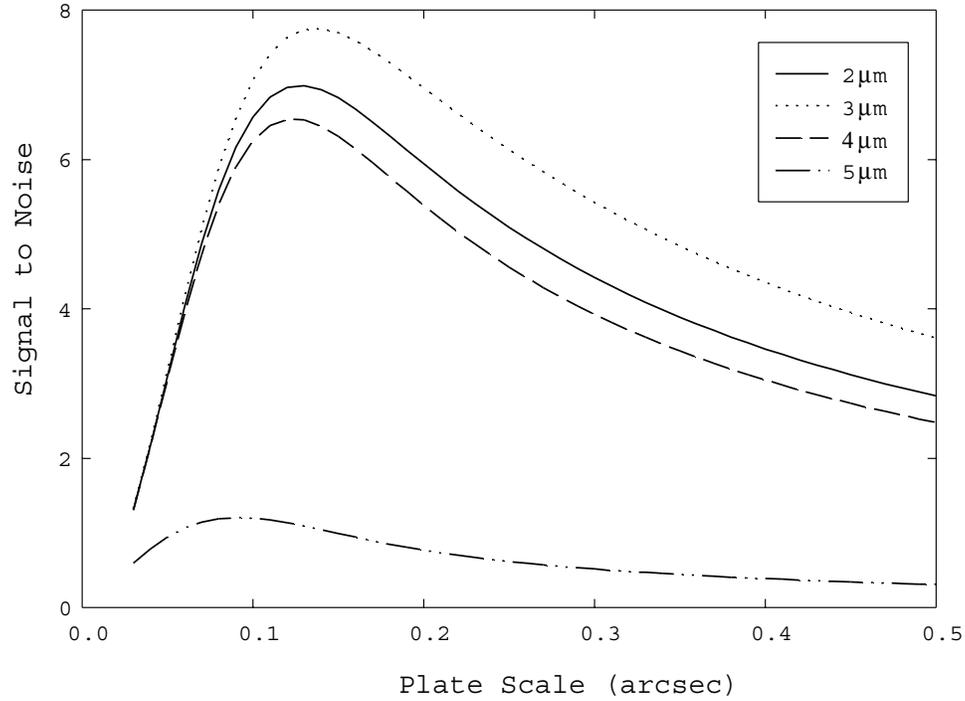

Figure 9. The signal to noise ratio vs. pixel size to detect an AB=28 mag galaxy in $10^5$ seconds of integration time at various wavelengths. We assume that the objects have exponential profiles with $r_{hl}$=0.1". At 5 mm, the zodiacal background is the highest and therefore the signal to noise ratio is the lowest and the peak occurs at smaller pixel scales.

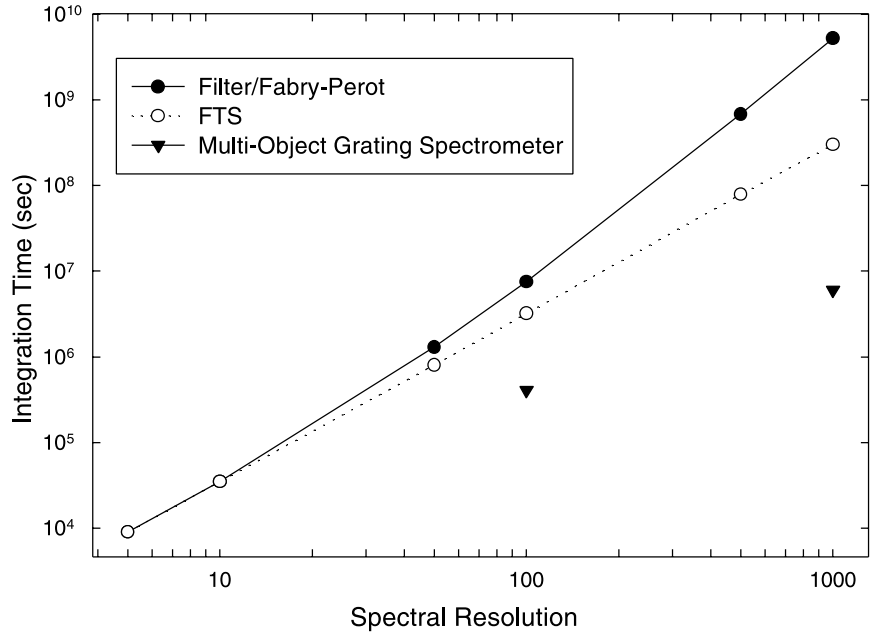

Figure 10. Total integration time required to obtain 1-5 µm spectrum ($F_{3\mu m}$ =10 nJy (AB=28.9 mag) at 10 σ over a 3' x 3' FOV. At low spectral resolution, filters and an FTS take the same amount of time to obtain a spectral data cube of the entire FOV. At R=100, the multi-object spectrometer can obtain spectra of all 4,000 galaxies at once and therefore takes the least time. Overhead in initial imaging observations and target preselection for the MOS observations are not included.